\documentclass[aip, reprint, superscriptaddress]{revtex4-1}
\usepackage{amsfonts, amssymb, amsmath, amsthm}
\usepackage{graphicx, epsfig, epsf, epstopdf}
\usepackage{multirow}
\usepackage{mathrsfs, bm, times}
\usepackage[table,xcdraw]{xcolor}
\usepackage[a4paper,colorlinks=true,
linkcolor=blue,citecolor=blue,
pdfauthor={ },
pdftitle={ },
pdfsubject={ },
pdfkeywords={ }]{hyperref}

\begin{document}

\title{Realization of programmable nanomechanical lattice with both nearest-neighboring and next-nearest-neighboring couplings}

\author{Shaochun Lin}
\affiliation{CAS Key Laboratory of Microscale Magnetic Resonance and Department of Modern Physics,
	University of Science and Technology of China, Hefei 230026, China}
\affiliation{Synergetic Innovation Center of Quantum Information and Quantum Physics,
	University of Science and Technology of China, Hefei 230026, China}

\author{Tian Tian}
\affiliation{CAS Key Laboratory of Microscale Magnetic Resonance and Department of Modern Physics,
	University of Science and Technology of China, Hefei 230026, China}
\affiliation{Synergetic Innovation Center of Quantum Information and Quantum Physics,
	University of Science and Technology of China, Hefei 230026, China}

\author{Pu Huang}
\affiliation{National Laboratory of Solid State Microstructures and Department of Physics, Nanjing University,
Nanjing 210093, China}

\author{Peiran Yin}
\affiliation{CAS Key Laboratory of Microscale Magnetic Resonance and Department of Modern Physics,
	University of Science and Technology of China, Hefei 230026, China}
\affiliation{Synergetic Innovation Center of Quantum Information and Quantum Physics,
	University of Science and Technology of China, Hefei 230026, China}

\author{Liang Zhang}
\affiliation{CAS Key Laboratory of Microscale Magnetic Resonance and Department of Modern Physics,
	University of Science and Technology of China, Hefei 230026, China}
\affiliation{Synergetic Innovation Center of Quantum Information and Quantum Physics,
	University of Science and Technology of China, Hefei 230026, China}

\author{Jiangfeng Du}
\thanks{Corresponding author: \href{mailto:djf@ustc.edu.cn}{djf@ustc.edu.cn}}
\affiliation{CAS Key Laboratory of Microscale Magnetic Resonance and Department of Modern Physics,
	University of Science and Technology of China, Hefei 230026, China}
\affiliation{Synergetic Innovation Center of Quantum Information and Quantum Physics,
	University of Science and Technology of China, Hefei 230026, China}
\affiliation{Hefei National Laboratory for Physical Sciences at the Microscale,
	University of Science and Technology of China, Hefei 230026, China}

\date{\today}

\begin{abstract}
The programmable artificial lattice, based on the controllability of coupling strengths and the scalability of multiple sites, is desperately desired in engineering metamaterials and exploring fundamental physics. In this work, we experimentally present a programmable lattice consisting of multiple paralleled nanomechanical resonators, whose internal interactions can be linearly manipulated by external voltages. Flexural modes of nearest-neighboring (NN) and next-nearest-neighboring (NNN) resonators are parametrically coupled through modulated electrostatic interactions. Particularly, in a wide range up to deep strong coupling regime, both the NN and NNN coupling strengths are precisely proportional to manipulation voltage. The realization of long-range coupling provides a promising prospect in constructing complex lattice structure, which is essential in investigating mechanical logic devices, topological physics and coherent phononic dynamics.
\end{abstract}

\maketitle
Progresses in artificial lattice have been persistently proceeded in various varieties of systems, including cold atoms\cite{Jaksch1998,Bloch2005,Eckardt2017}, optical waveguides\cite{Malkova2009,Garanovich2012,Blanco-Redondo2016}, surface acoustic waves\cite{He2018,Raguin2019}, phononic crystals\cite{Zhu2017,Cha2018} and nanomechanical resonators\cite{Mahboob2013,Huang2016,Gajo2017}. Recent works have realized the lattices as functional devices\cite{Cencillo-Abad2016,Suesstrunk2017}, complex networks\cite{Fon2017, Matheny2019}, and topological structures\cite{King2018, Kempkes2019,Tian2019}.
However, deficiencies of such systems appear when it comes to adiabatic dynamics\cite{Ke2016Topological,Rosa2019Edge} and non-equilibrium physics\cite{Heyl2018Dynamical,Swingle2018Unscrambling}, where individually tunable couplings are required.
Therefore,  the programmable lattice, possessing completely controllable coupling between multiple sites, has attracted considerable attention in quantum simulation\cite{Kempkes2019}, metamaterial\cite{Cha2018a}, topological physics\cite{Hu2019}, and information processes\cite{Luo2018}.

Recent years have witnessed the promising potential towards programmable lattice in nanomechanical systems\cite{Okamoto2016,Suesstrunk2015Observation,SerraGarcia2018Observation}. Heretofore, tunable couplings between two resonators have been accomplished\cite{Sulkko2010,Faust2013Coherent,Doster2019Collective}. However, since the conventional coupling mechanism is based on strain force\cite{Bueckle2018Stress,Faust2012Nonadiabatic}, which is rigid due to intrinsically fixed geometry, especially for multiple resonators, therefore, a practical approach towards programmable nanomechanical lattice still remains challenge.

In this work, we provide a practical programmable nanomechanical lattice employing electrostatic parametric couplings between resonators instead of the strain couplings. The parametric couplings are induced when we modulate the applied external manipulation voltages at the frequency-difference between the flexural modes of target resonators. Under such mechanism, the coupling strength is precisely proportional to the manipulation voltage. The NN couplings possess a wide tunable range up to $13$ times the decay rates, meanwhile the NNN coupling can also reach $7$ times the decay rates. Moreover, we demonstrate the programmability via constructing NN coupled lattice and NNN coupled lattice on the same nanomechanical device, respectively.

\begin{figure}
	\centering
	\includegraphics[width=1\linewidth]{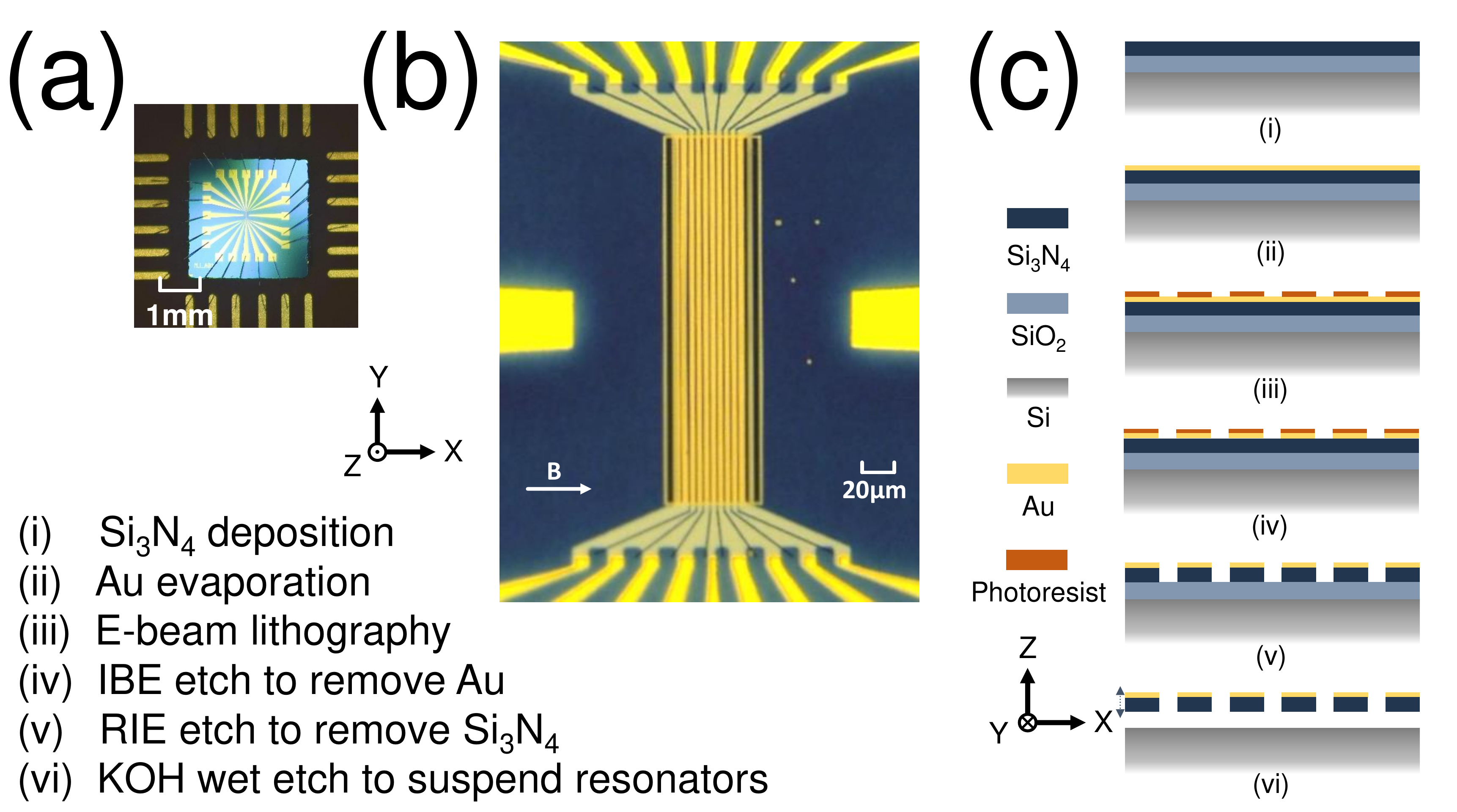}
	\caption{\label{fig1:device}
	Overview of the device.
(a) Photograph of the on-chip device and wire-bonded chip-carrier. The size of the chip is approximately $3~\rm mm \times 3~\rm mm$.
(b) Microscope photograph of $11$ closely adjacent doubly-clamped silicon nitride string resonators, in the size of $200~\rm \mu m$ long, $3~\rm \mu m$ width and $100~\rm nm$ thick, with $15~\rm nm$ Au coated. The gap between NN resonators is $500~\rm nm$. Those resonators vibrate along Z direction, when the magnetic field is applied on X axis.
(c) Nano-fabrication processes of Au-coated silicon nitride string resonators array.
	}
\end{figure}

This on-chip device consists of $11$ silicon nitride string resonators at the size of $200~\rm \mu m$ long, $3~\rm \mu m$ wide, and $100~\rm nm$ thick. They are closely adjacent within a $500~\rm nm$ gap and evaporated with a $15~\rm nm$ thick Au layer so as to connected with outside electrodes. As shown in Fig.~\ref{fig1:device}(a), the electrodes are wired-bonded to outside chip-carrier. The string resonators are driven via voltage applying on the Au layer following standard magneto-motive technique\cite{Cleland1999External} under 1T magnetic field along X direction, shown in Fig.~\ref{fig1:device}(b). The first flexural modes of these resonators, along Z direction, are utilized in this experiment.

Fig.~\ref{fig1:device}(c) illustrates the main processes of device fabrication. (i) At the beginning, a $100~\rm nm$ thick layer of silicon nitride is deposited by low pressure chemical vapor deposition (LPCVD). (ii) Then we evaporate $15~\rm nm$ Au layer on the top of silicon nitride. (iii) By means of e-beam lithography, the shape of string resonators is patterned and protected by photoresist. (iv) and (v) we etch the extra Au and silicon nitride in the gap between resonators by ion beam etching (IBE) and reactive ion etching (RIE), respectively. (vi) Finally, the resonators array is suspended after potassium hydroxide (KOH) wet etching. Here, to be noticed, on the purpose of strong coupling, the gap between resonators should be as close as possible to enhance the electrostatic field intensity (discussed later). However, those long and closely adjacent string resonators tend to bend and contact with each other after suspension, especially for multiple resonators. To solve this technical difficulty, we gradually replace the environment with low surface tension surroundings while suspending (liquid transferring from KOH to the deionized water, then to the isopropanol), so that the surface force would not attract those string resonators. Moreover, it has to be emphasized that, so as to remove the intrinsic strain coupling, the connected $\rm Si_{3}N_{4}$ substrate is cut off by focused ion beam etch. In this case, electrostatic coupling is dominant between resonators.

\begin{figure}
	\centering
	\includegraphics[width=1\linewidth]{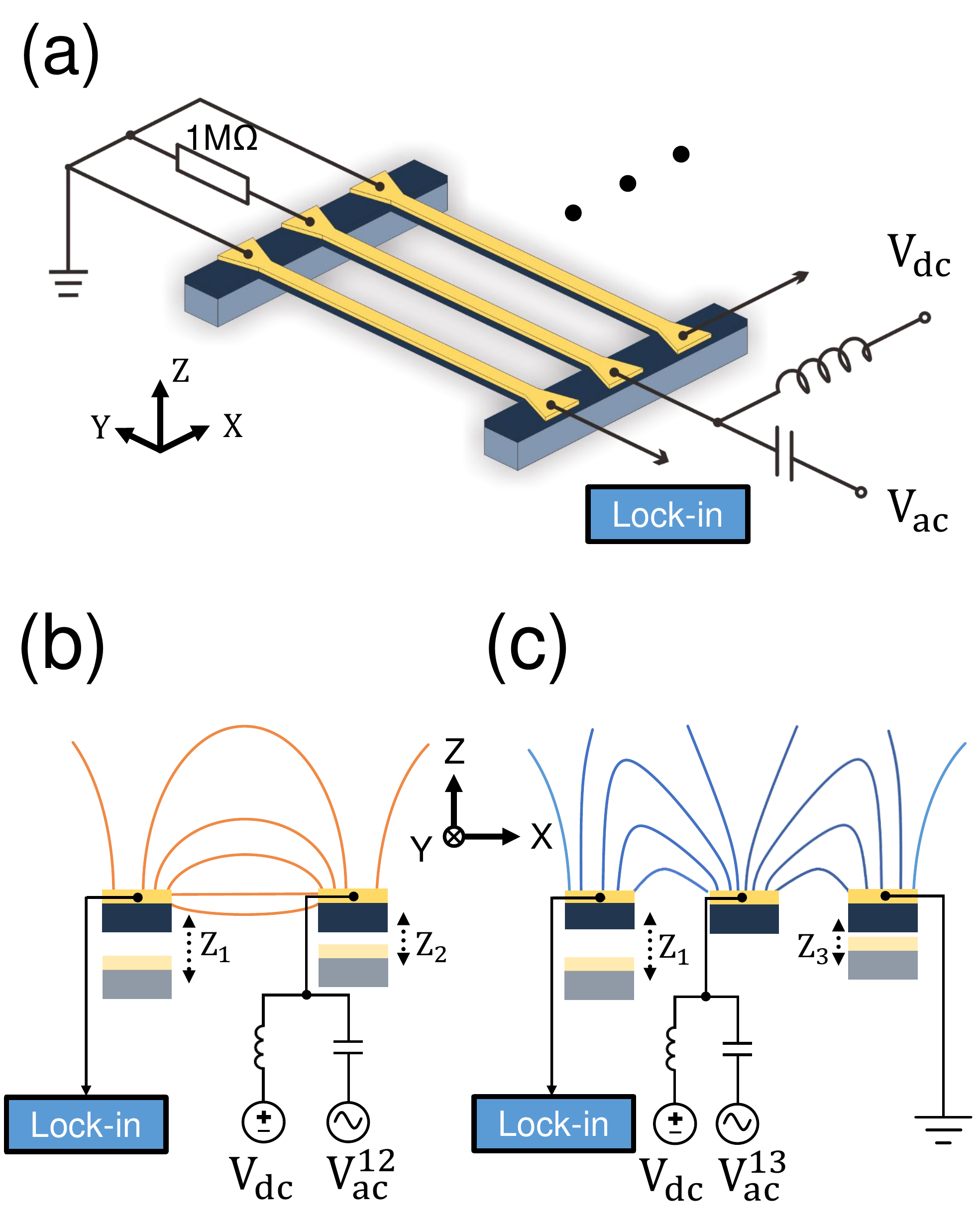}
	\caption{
		\label{fig2:mechanism}
	Coupling mechanism of programmable lattice dominated by electrostatic interaction.
	(a) The schematic of $\rm 3$ resonators in this array whose inside coupling strength is completely controlled by outside manipulation voltage. Here, $\rm V_{dc}\gg\rm V_{ac}$ (manipulation~voltage) to reduce the frequency shift induced by $\rm V_{ac}$.
	(b) and (c) The schematics of NN and NNN couplings, respectively. The $V_{\mathrm{ac}}^{12}$ and $V_{\mathrm{ac}}^{13}$ are applied at the frequencies of $|\omega_2-\omega_1|$, $|\omega_3-\omega_1|$.
    All the lock-in amplifiers are used to excite and measure the vibration of resonators.
	}
\end{figure}

Fig.~\ref{fig2:mechanism} demonstrates the principle of NN and NNN couplings. Fig.~\ref{fig2:mechanism}(a) displays the schematic of programmable lattice whose inside coupling strength is completely controlled by outside manipulation voltage ($V_\mathrm{ac}$). The device is settled in the environment of 77K (liquid nitrogen cooling) and high vacuum in order to keep temperature stable and dissipation low. Each Au coated resonator connects with its own electrode and bond to external circuit through Al wire, so that each of them can be manipulated individually. In the experiments, resonators work in linear regime, which is guaranteed by weak exciting power. As Fig.~\ref{fig2:mechanism}(a) shows, manipulation voltages combined with dc voltages are applied via bias-tees. Here, we set $ V_{\mathrm{dc}}\gg\rm V_{\mathrm{ac}}$, in order to reduce the frequency shift of the resonator caused by $V_{\mathrm{ac}}$. Fig.~\ref{fig2:mechanism}(b) and (c)  diagrams the inside coupling between two sites responding to manipulation voltage.

Employing standard magneto-motive procedure, we acquire the fundamental eigenfrequencies of all the resonators in this system, around $2\pi \times 898~\rm{KHz}$. Meanwhile, the Au layer coated on the high-stressed silicon nitride resonators certainly decreases the Q-factors\cite{Verbridge2006}, but it can still reach over $1\times 10^5$ under the fabrication processes. All the measurement is taken in the environment of $77~\rm K$ and $3 \times 10^{-6}~\mathrm{Pa}$. Here, we employ high-Q resonators to suppress the incoherent strain couplings. In addition, high-Q resonators bring benefits in longer coherence time and better signal-to-noise ratio, which make it possible to observe the dynamics of individual sites.

For NN coupling, as shown in Fig.~\ref{fig2:mechanism}(b), out-of-plane modes are coupled through electric field generated by applying $V_{\rm p}(t)=V_{\rm dc}+V_{\rm ac}\cos(\omega_{\rm p} t)$ between each two resonators.
The resulting electrostatic force is proportional to the out-of-plane field strength $E_z$ and the induced charge $Q$ on grounded resonator\cite{Unterreithmeier2009}:
\begin{align}
F_z(t) =\int QE_z \mathrm{d} x \mathrm{d} y \propto V_{\rm p}^2(t).
\end{align}
Usually, the bias dc voltage $V_{\rm dc}$ is fixed at an appropriate value no more than 10$\rm V$, while the manipulation voltage $V_{\rm ac}$ is much smaller than $V_{\rm dc}$. Neglecting $V_{\rm ac}^2$ elements, the modulated force leads to $F_z\approx c_{12}(z)V^2_{\rm dc}+2c_{12}(z)V_{\rm dc}V_{\rm ac}\cos(\omega_{\rm p} t)$, with $c_{12}(z)$ being a coefficient depending on the relative position of the coupled resonators $z=z_1-z_2$.
This modulated force to the first order in $z$ yields:
\begin{align}
F_z = F_0 + \frac{\partial c_{12}}{\partial z}z [V_{\rm dc}^2+2V_{\rm dc}V_{\rm ac}\cos(\omega_{\rm p}t)].
\end{align}
The static term $F_0$ results in a new equilibrium position and can be ignored. However, the relative-position-dependent terms lead to shifted resonance frequencies and effective coupling. The motion equations of two coupled resonators are described as:
\begin{align}
\label{eq3:coupling}
\begin{cases}
&\ddot{z}_{1}+\gamma_{1} \dot{z}_{1}+\tilde{\omega}_{1}^2 z_1 \approx 2c'_{12}(0)V_{\rm dc}V_{\rm ac}\cos(\omega_{\rm p} t)z_{2}, \\
&\ddot{z}_2+\gamma_2 \dot{z}_1+\tilde{\omega}_2^2 z_1 \approx2c'_{12}(0)V_{\rm dc}V_{\rm ac}\cos(\omega_{\rm p} t)z_1.
\end{cases}
\end{align}
Here $\gamma_{\mathrm{1,2}}$ are the decay rates. Resulting from \eqref{eq3:coupling}, $\tilde{\omega}_{1,2} = \sqrt{\omega^2_{1,2}-c'_{12}(0)\frac{V_{\rm dc}^2}{m}}$ indicate the shifted eigenfrequencies of the two fundamental modes and $c'_{12}(0)=\frac{\partial c_{12}}{\partial z}\big|_{z=0}$ is the effective interaction coefficient between the adjacent resonators at the new equilibrium position.
The NN resonators are parametric coupled when the pump frequency $\omega_{\rm p}$ comes to the difference of shifted eigenfrequencies $\omega_{\rm p}=|\tilde{\omega}_1-\tilde{\omega}_2|$.
The effective coupling strength $g_{12}\equiv c'_{12}(0)V_{\rm dc}V_{\rm ac}$, is proportional to the manipulation voltage $V_{\rm ac}$.

It should be pointed out that, the interaction coefficients $c_{12}(z)$ for different pairs of adjacent resonators are unequal, due to the minor differences from the fabrication process and electric field distribution, such as the differences of beam widths, the gap distances and so on. In this work, we demonstrate the largest coupling strength that can be achieved on a single lattice. Meanwhile, we demonstrate the NNN coupling scheme on the same device.

Similarly, the direct NNN coupling between separated resonators is induced by the applied manipulation voltage on the middle one, as the schematic diagram shows in Fig.~\ref{fig2:mechanism}(c). The vibration of the 1st resonator (left) leads to a periodic change of electric field acting on the 3rd resonator (right). The resulting electrostatic force is dependent on the relative position of the non-adjacent resonators, even with the 2nd resonator (middle) staying at rest, $z_2=0$. Thus, the equations describing the motion of NNN coupled resonators are similar as equations \eqref{eq3:coupling}, replacing the position-dependent coefficient $c_{12}(z\equiv z_2-z_1)$ by another one, denoted by $c_{13}(z'\equiv z_3-z_1)$,
\begin{align}
\label{eq4:next-nearest-coupling}
\begin{cases}
&\ddot{z}_1+\gamma _1\dot{z}_1+\tilde{\omega}_{1}^2 z_1= 2c'_{13}(0)V_{\rm dc}V_{\rm ac}\cos(\omega_{\rm p}t)z_3, \\
&\ddot{z}_3+\gamma_3\dot{z}_3+\tilde{\omega}_{3}^2z_3= 2c'_{13}(0)V_{\rm dc}V_{\rm ac}\cos(\omega_{\rm p}t)z_1,
\end{cases}
\end{align}
where $c'_{13}(0)=\frac{\partial c_{13}}{\partial z'}\big|_{z'=0}$. Equations \eqref{eq4:next-nearest-coupling} indicate that for NNN coupling, the coupling strength $g_{13}\equiv c'_{13}(0)V_{\mathrm{dc}}V_{\mathrm{ac}}$ still stays in linear regime with the manipulation voltage. Generally, $g_{13}$ is less than NN coupling $g_{12}$, considering the weaker position-dependent coefficient $c'_{13}(0)<c'_{12}(0)$.

\begin{figure}
	\centering
	\includegraphics[width=1\linewidth]{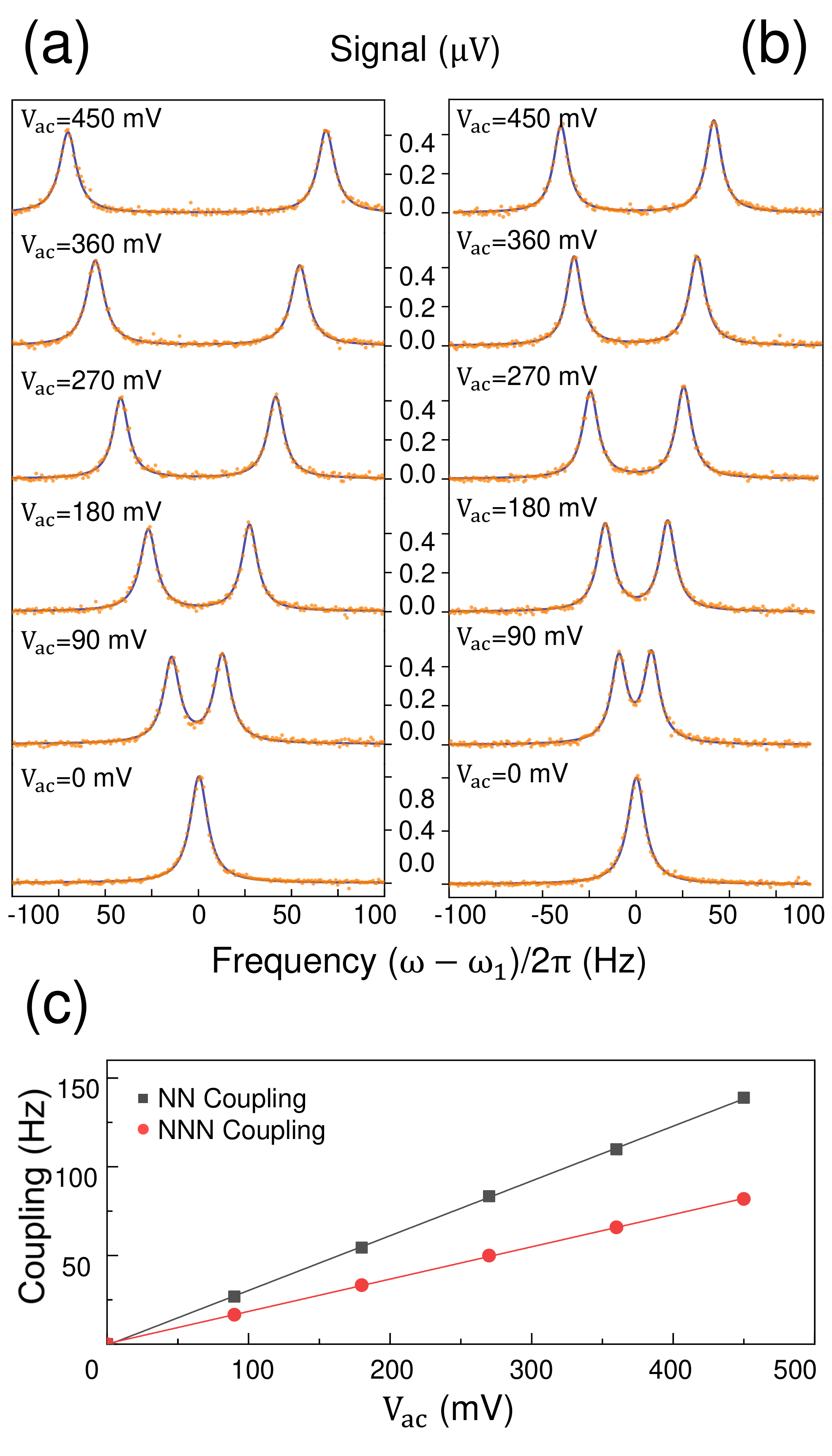}
	\caption{\label{fig3:linear}
		The experimental results of NN and NNN couplings.
		(a) and (b) The response spectra of resonator 1 under different manipulation voltages ($V_{\rm ac}$) for NN and NNN couplings, respectively. The peak splittings denote the coupling strength.
		(c) Compare of NN and NNN couplings under different manipulation voltages.
	}
\end{figure}

According to the coupling mechanism and experimental measurement, Fig.~\ref{fig3:linear} displays the linear relationship between coupling strength (represented by frequency split) and the manipulation voltage $ V_{\rm ac}$, for both NN and NNN in this lattice. The NN coupling, taking the first and second resonators for example, exhibits an excellent linear relationship with manipulation voltage $V_{\rm ac}$, as shown in Fig.~\ref{fig3:linear}(a). During the experiment, dc voltage keeps at $5~\rm V$ and combines with ac voltage via bias-tee at $\omega_{\rm p}=\omega_{2}-\omega_{1}$, applying on the second resonator, shown in Fig.~\ref{fig2:mechanism}(b). The spectral split, standing for coupling strength, broadens linearly while $V_{\rm ac}$ increases from $0$ to $450~\rm mV$, which is monitored on the first resonator by lock-in amplifier. In this way, the coupling strength $g_{12}$ from $0$ to $2\pi\times 140\rm{Hz}$ ($13\gamma_1$) can be implied. Analogously, the NNN coupling between the first and third resonators, illustrated in Fig.~\ref{fig3:linear}(b), is measured on the first resonator, while modulation voltage with $\omega_{\rm p}=\omega_{3}-\omega_{1}$ is applying on the second resonator, and the third resonator are grounded, see Fig.~\ref{fig2:mechanism}(c). The experimental results indicate that the NNN coupling  $g_{13}$  is relatively weaker than the NN coupling under the same $V_{\rm dc}$ and $V_{\rm ac}$, which is still able to tune from $0$ to $2\pi\times82\rm Hz$ ($7\gamma_1$), shown in Fig.~\ref{fig3:linear}(b-c). These wide tunable range and linear relationships produce precisely programmability for us to control the interaction between sites by manipulating voltages outside.

\begin{figure}
	\centering
	\includegraphics[width=1\linewidth]{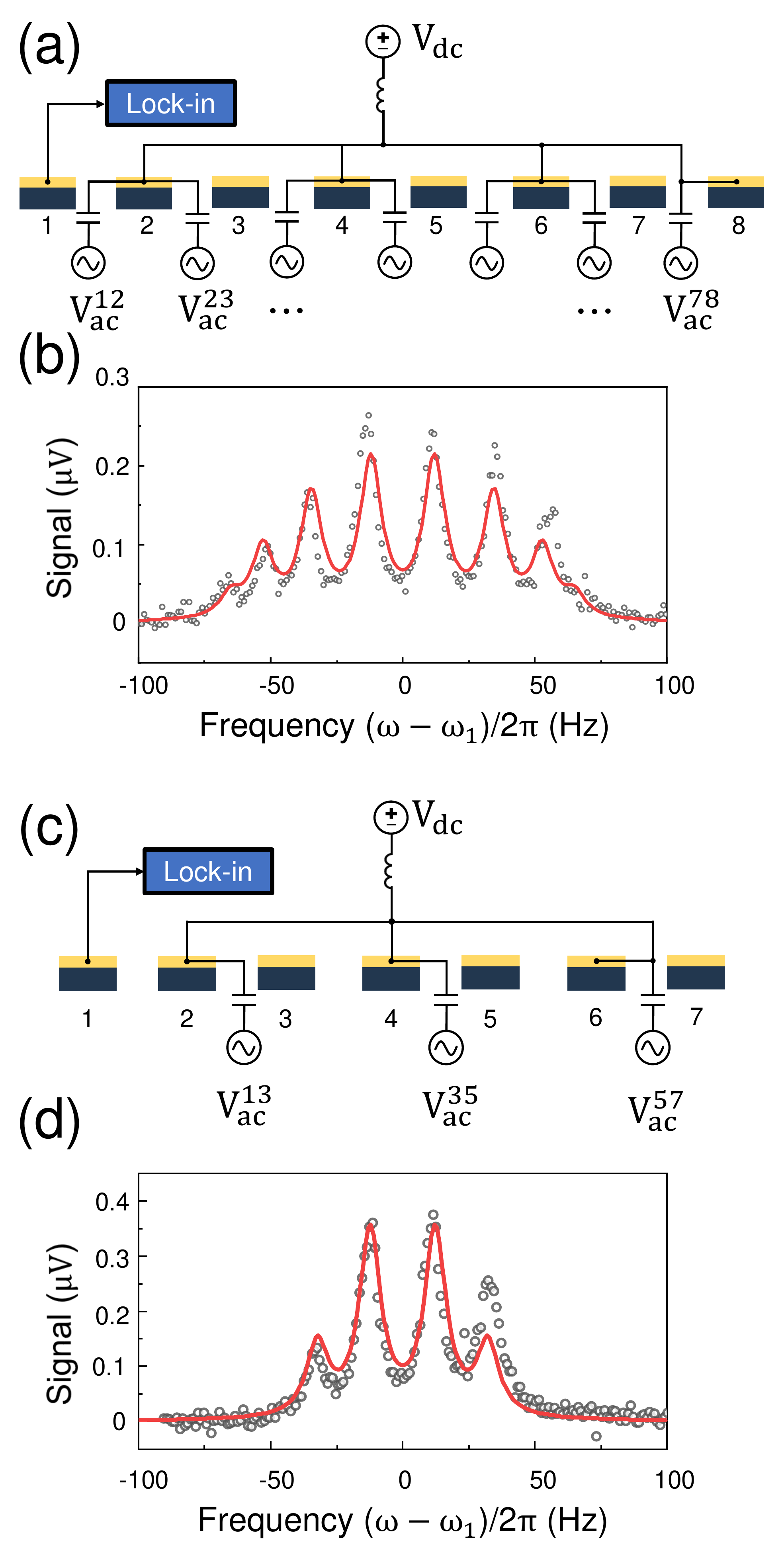}
	\caption{\label{fig4:lattice}
		(a)(c) Realization of a NN coupled and NNN coupled lattices, respectively.
		(b)(d) Response spectrum of the NN coupled and NNN coupled lattice. The circles illustrate the experimental data and the lines stand for the numerical results.
	}
\end{figure}

To exhibit the electrostatic coupling mechanism in the entire array, here, we deliver the different lattices on the same device via various manipulation voltages.
In experiment, 8 sites of the resonators array are NN coupled via tuning the manipulation voltages $V_{\rm ac}^{i,i+1}$ at the frequencies of $|\omega_{i+1}-\omega_{i}|$ on even-numbered resonators, as Fig.~\ref{fig4:lattice}(a) shows. All the NN coupling strengths are set to $2\pi\times 70~{\rm Hz}$ ($V_{\mathrm{dc}}=\rm 5~\rm V$). In this way, we construct a entirely NN coupled lattice, whose response spectrum theoretically indicates 8 separated eigenmodes. We experimentally measure the spectrum by exciting and reading the vibration amplitude of the edge resonator in this lattice, utilizing a lock-in amplifier. The 8 peaks in the spectrum correspond to the eigenmodes, shown in Fig.~\ref{fig4:lattice}(b). There is a slightly mismatch between experimental and theoretical results due to the frequency shifts caused by manipulation voltages, which can be reduced by increasing $V_{\rm dc}$ and decreasing $V_{\rm ac}$.  Analogously, Fig.~\ref{fig4:lattice}(c) indicates that we construct a entirely NNN coupled lattice on the same device. The $V_{\rm ac}^{i, i+2}$ on the even numbered resonators are tuned at the frequencies of $|\omega_{i+2}-\omega_{i}|$, while $V_{\rm dc}$ keeps at 5V. This NNN couplings are also set to $2\pi\times 70~{\rm Hz}$. Fig.~\ref{fig4:lattice}(d) exhibits the experimental response spectrum with 4 peaks.

In conclusion, we realize a programmable nanomechanical lattice on multiple closely adjacent resonators with both NN and NNN couplings.  Both  couplings can be linearly controlled with a wide tunable range, determined by the electrostatic parametric coupling mechanism. The unconventional coupling mechanism makes up for the weakness of strain coupling in lacking programmability and scalability. The programmability is demonstrated through constructing entirely NN and NNN coupled lattices. Meanwhile, the fabrication of this device permits well-scalability, able to reach tens of resonators.
This programmable nanomechanical lattice possesses promising potential in the fields of mechanical functional devices\cite{Brandenbourger2019Non,Nassar2017Non}, topological dynamics\cite{Weidinger2017Dynamical,Heyl2018Detecting}, and information processing\cite{Lee2017Topical,Bienfait2019}. Moreover, the long-range couplings, represented by NNN couplings,  blaze a path towards high-dimensional structures\cite{Yuan2018}.
\begin{acknowledgements}
		This work was supported by the National Key R\&D Program of China (Grant No. 2018YFA0306600), the CAS (Grants No. GJJSTD20170001 and No. QYZDY-SSW-SLH004), and Anhui Initiative in Quantum Information Technologies (Grant No. AHY050000). This work was partially carried out at the USTC Center for Micro and Nanoscale Research and Fabrication. We thank Zehua Tian for his valuable suggestions.
\end{acknowledgements}

\section*{Data Availability}
The data that support this work are available from the corresponding author upon reasonable request.
\bibliography{programmable}
\end{document}